\documentclass[prd,tightenlines,showpacs,nofootinbib,twocolumn]{revtex4-1}

\usepackage[english]{babel}

\usepackage[utf8x]{inputenc}
\usepackage[T1]{fontenc}
\usepackage{ucs}
\usepackage{textcomp}
\usepackage{microtype}
\usepackage{lmodern}
\usepackage{pdfpages}

\usepackage{url}
\usepackage{hyperref}

\usepackage{xspace}
\usepackage{lastpage}
\usepackage{stmaryrd}

\usepackage{graphicx}
\usepackage{rotating}
\usepackage{empheq}
\usepackage{ifthen}
\usepackage{tensor}
\usepackage{paralist}
\usepackage{wrapfig}
\usepackage{array}

\usepackage{amsmath}
\usepackage{amsfonts}
\usepackage{amssymb}
\usepackage{mathrsfs}
\usepackage{extarrows}
\usepackage{verbatim}
\usepackage{upgreek}

\usepackage{natbib}

\newcommand\define{\equiv}

\newcommand\vect[1]{\boldsymbol{#1}}

\newcommand\e[1]{_{\text{#1}}}

\newcommand\U[1]{\:\mathrm{#1}}

\newcommand{\dd}{\mathrm{d}}
\newcommand{\Dd}{\mathrm{D}}

\newcommand{\ddf}[3][]{\frac{\dd^{#1} #2}{\dd {#3}^{#1}}}
\newcommand{\Ddf}[3][]{\frac{\Dd^{#1} #2}{\dd {#3}^{#1}}}

\renewcommand\lim[2]{\underset{ #1 \rightarrow #2 }{ \mathrm{lim} } \,}

\newcommand{\delimiters}[4][]{
\ifthenelse{ \equal{#1}{1} }{  #2 #3 #4  }
					{ \ifthenelse{\equal{#1}{2}}{ \big#2 #3 \big#4 }
						{ \ifthenelse{\equal{#1}{3}}{ \Big#2 #3 \Big#4 }
							{ \ifthenelse{\equal{#1}{4}}{ \bigg#2 #3 \bigg#4 }
								{ \ifthenelse{\equal{#1}{5}}{ \Bigg#2 #3 \Bigg#4 }
									{ \left#2 #3 \right#4 }
								}
							}
						}
					}
													}

\newcommand{\pa}[2][]{\delimiters[#1]{(}{#2}{)}}

\newcommand{\source}{\text{s}}
\newcommand{\obs}{\text{o}}

\newcommand{\tidal}{\mathcal{R}}

\newcommand{\wl}{\mathscr{L}}

\newlength{\boxtitlelength}
\newlength{\halfrulelength}
\newcommand{\boxtitle}[1]{\footnotesize\bf{\:#1\:}}

\definecolor{blue4}{RGB}{0,0,143}
\definecolor{red4}{RGB}{143,0,0}
\definecolor{orange}{RGB}{255,128,0}
\definecolor{darkcyan}{RGB}{0,128,128}
\definecolor{olive}{RGB}{0,128,0}
\definecolor{purple}{RGB}{128,0,128}
\definecolor{cyan2}{RGB}{0,255,255}
\definecolor{fushia}{RGB}{255,0,255}
\definecolor{mygray}{gray}{0.5}
\definecolor{lightgray}{gray}{0.85}

\newcommand{\araa}{ARAA}
\newcommand{\pasp}{PASP}
\newcommand{\jcap}{JCAP}

\newcommand{\UR}{^{\rm part}}
\newcommand{\REF}{^{\rm ref}}
\newcommand{\FGMV}{FGMV15\xspace}

\begin{document}

\title{On the time delay between ultra-relativistic particles}

\author{Pierre Fleury}
\email{pierre.fleury@uct.ac.za}
 
\affiliation{
Department of Mathematics and Applied Mathematics, University of Cape Town,
Rondebosch 7701, Cape Town, South Africa,\\
Department of Physics, University of the Western Cape,
Robert Sobukwe Road, Bellville 7535, South Africa.
}

\begin{abstract}
The time delay between the receptions of ultra-relativistic particles emitted simultaneously is a useful observable for both fundamental physics and cosmology. The expression of the delay when the particles travel through an arbitrary spacetime has been derived recently by Fanizza et al., using a particular coordinate system and self-consistent assumptions. The present article shows that this formula enjoys a simple physical interpretation: the relative velocity between two ultra-relativistic particles is constant. This result reveals an interesting kinematical property of general relativity, namely that the tidal forces experienced by ultra-relativistic particles in the direction of their motion are much smaller than those experienced orthogonally to their motion.
\end{abstract}

\date{\today}
\maketitle

%%%%%%%%%%%%%%%%%%%%%%%%%%%%%%%%%%%%%%%%%%%%%%%%%%%%

\section{Introduction}

The time delay between the reception of ultra-relativistic (UR) particles has historically been proposed in 1968 as an astrophysical observable to measure the mass of the electronic neutrino~\cite{1968JETPL...8..205Z}. This method has notably been applied to the delay between photons and neutrinos emitted during the supernova explosion~SN1987A, yielding the upper limit~$m_\nu\leq16\U{eV}$ for the mass of the electronic neutrino (see e.g. ref.~\cite{1989ARA&A..27..629A} and references therein). Although much less constraining than today's limits on neutrino masses obtained by the \textit{Planck} mission~\cite{2015arXiv150201589P}, the photon-neutrino delay observed with SN1987A has been nevertheless one of the main arguments against the OPERA erroneous measurement of superluminal neutrinos~\cite{2012JHEP...10..093A}.

The idea of using time delays between UR particles as a cosmological probe is more recent~\cite{2000PhLB..473...61S} (see also the short review~\cite{2001neph.conf...14S}). Though observational applications still have to face technical difficulties~\cite{2014PASP..126..885G}, the time delays between, e.g., cosmic rays and $\gamma$-ray bursts are expected to provide independent measurements of the cosmological parameters in the future. However, even with perfect sources and instruments, an irreducible uncertainty comes from the fact that the particles propagate in a locally inhomogeneous universe, and are therefore affected by gravitational phenomena. This issue was recently tackled by the authors of ref.~\cite{2016PhLB..757..505F}, hereafter \FGMV, who derived a general expression for the time delay within an arbitrary spacetime, generalizing the formula proposed in ref.~\cite{2000PhLB..473...61S} for the Friedmann-Lema\^itre-Robertson-Walker universe.

The result of \FGMV is the following. Consider two particles $P_1$, $P_2$ emitted at the same event $S$ with different energies. Since one of them is slightly slower than the other, those particles are received at different times $t_1, t_2$ by the observer, whose difference is
\begin{equation}\label{eq:time_delay}
\Delta t \define t_2 - t_1 = \pa{ \frac{m_2^2}{2 E_2^2} - \frac{m^2_1}{2 E_1^2} } \int_{t\e{s}}^{t_1} \frac{\dd t}{1+z(t)},
\end{equation}
at lowest order in the inverse of the gamma factors $\gamma_i\define E_i/m_i \gg 1$, where $m_i$ and $E_i$ are respectively the rest mass and the energy of $P_i$, as measured at reception in the observer's frame. In eq.~\eqref{eq:time_delay}, the redshift~$z$ is not necessarily cosmological, because the formula is valid for any geometry, but it rather relies on an arbitrary 3+1 foliation of spacetime such that the coordinate $t$ coincides with the observer's proper time. The integral over $t$ must be understood as an integral along the worldline of $P_1$, which is approximately a null geodesic. It is remarkable that eq.~\eqref{eq:time_delay} has exactly the same form as in a strictly homogeneous and isotropic universe.

In \FGMV, this result was derived using the geodesic-light-cone formalism~\cite{2011JCAP...07..008G,2016JCAP...06..008F}, and relying on a self-consistent ansatz. However, as mentioned by the authors themselves, the simplicity of eq.~\eqref{eq:time_delay} suggests the existence of a more general derivation, which is precisely the purpose of the present paper. In sec.~\ref{sec:interpretation}, I show that the time-delay formula is actually equivalent to assuming that the relative velocity of two UR particles is constant during their travel. I then physically justify this surprising assumption in sec.~\ref{sec:tidal_forces_UR}, which reveals a general mechanism about how UR particles experience tidal forces.

%%%%%%%%%%%%%%%%%%%%%%%%%%%%
\section{Physical interpretation of the time-delay formula}
\label{sec:interpretation}
%%%%%%%%%%%%%%%%%%%%%%%%%%%%

Equation~\eqref{eq:time_delay} has the advantages of directly involving observable quantities, and exhibits a dependence in the cosmological parameters via~$z$. However, its physical meaning is hidden and therefore requires some reformulation. First notice that the prefactor of the integral corresponds to the difference between the velocities~$v_i$ of the particles as measured by the observer:
\begin{align}
v_1 - v_2 &= \sqrt{1-\frac{1}{\gamma_1^2}} - \sqrt{1-\frac{1}{\gamma_2^2}} \\
				&= \frac{1}{2\gamma_2^2} - \frac{1}{2\gamma^2_1} + \mathcal{O}(\gamma^{-4}).
\end{align}

Second, the integral of eq.~\eqref{eq:time_delay} is proportional to the proper travel time~$\tau_1$ of $P_1$ from its emission at $S$ to its observation at $O_1$. Indeed, since the particle is UR the evolution of its energy is essentially encoded in the lightlike redshift~$z$ as
\begin{equation}
1+z(t) \approx \frac{E(t)}{E_\obs} = \frac{\gamma(t)}{\gamma_\obs} = \frac{1}{\gamma_\obs} \ddf{t}{\tau},
\end{equation}
where a subscript o indicates the observed value of a quantity, and $\tau$ denotes the proper time of $P_1$; whence
\begin{equation}
\int_{t\e{s}}^{t_1} \frac{\dd t}{1+z(t)} \approx \gamma_1 \tau_1,
\end{equation}
so that eq.~\eqref{eq:time_delay} takes the form
\begin{equation}\label{eq:time_delay_reformulation_1}
\Delta t = \gamma_1 (v_1-v_2) \,\tau_1.
\end{equation}

Though simpler than the former, the latter formula involves quantities defined in different frames, which makes it hard to interpret. The last step thus consists in translating eq.~\eqref{eq:time_delay_reformulation_1} into a relation between quantities in $P_1$'s frame only. More specifically, we are going to relate the observed time delay $\Delta t$ to the distance~$\ell$ that separates the particles in $P_1$'s frame, when $P_1$ is detected by the observer.

\begin{figure}[h!]
\centering
\includegraphics[scale=1]{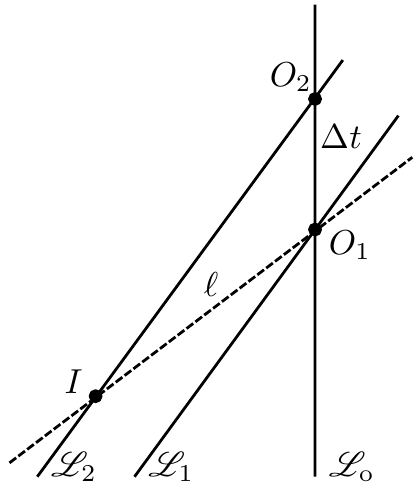}
\caption{Relation between (i) the time delay~$\Delta t$ of two UR particles following the worldlines $\wl_1$, $\wl_2$, in the frame of the observer~$\wl_\obs$; and (ii) the proper distance~$\ell$ between those particles in the frame of the fastest one when it is detected by the observer. Solid lines indicate worldlines, while the dashed line represents the set of all events which are simultaneous with $O_1$ in $P_1$'s frame.}
\label{fig:delay}
\end{figure}

The geometry of the problem is depicted in Fig.~\ref{fig:delay}. The worldline~$\wl_i$ of the particle $P_i$ intersects the observer's worldline~$\wl_\obs$ at the even~$O_i$. Define $I$ as the event of $\wl_2$ which is simultaneous with $O_1$ in the frame of $P_1$. This event therefore indicates the spatial position of $P_2$ in this frame when $P_1$ is detected by the observer. The spacetime separation~$\Delta s^2(I,O_1)$ is therefore equal to $\ell^2$.

In the following, it will be convenient to work with a Fermi normal coordinate system~\cite{2004rtmb.book.....P}~$(t,x^i)$ about $\wl_\obs$, so that spacetime appears nearly flat in the vicinity of this worldline. As a consequence, the geodesics $\wl_1, \wl_2$ are essentially straight lines in the domain of interest. If we choose the axes of $(x^i)$ so that $\vect{\partial}_1$ at $O_1$ is aligned with the direction of $O_1 E$, then the problem becomes spatially one-dimensional, and we only have to consider the coordinate~$x^1\define x$.

Because $I\in\wl_2$, its coordinates~$(t_I,x_I)$ satisfy
\begin{equation}\label{eq:E_in_L2}
x_I = v_2 (t_I-\Delta t).
\end{equation}
Besides, a simple Lorentz transformation relates those coordinates to their counterpart~$(0,-\ell)$ in $P_1$'s frame,
\begin{align}
0 &= \gamma_1 (t_I-v_1 x_I) \label{eq:Lorentz_time}\\
-\ell &= \gamma_1 (x_I - v_1 t_I).\label{eq:Lorentz_space}
\end{align}
Combining eqs.~\eqref{eq:E_in_L2}, \eqref{eq:Lorentz_time}, and \eqref{eq:Lorentz_space} then yields
\begin{equation}
\ell = \frac{1}{\gamma_1}\frac{v_2}{1 - v_1 v_2} \, \Delta t,
\end{equation}
which, once introduced in eq.~\eqref{eq:time_delay_reformulation_1}, finally gives
\begin{empheq}[box=\fbox]{equation}\label{eq:time_delay_meaning}
\ell = \frac{v_1 - v_2}{1 - v_1 v_2} \, \tau_1 = |v_{2/1}|\,\tau_1,
\end{empheq}
where we have recognised the expression of the velocity~$v_{2/1}$ of $P_2$ in $P_1$'s frame~\cite{GourgoulhonRR} at the observation event. Equation~\eqref{eq:time_delay_meaning} is equivalent to the time-delay formula~\eqref{eq:time_delay} at lowest order in the inverse gamma factors, but its meaning is clearer: since the distance~$\ell$ is proportional to the travel time~$\tau_1$, \emph{the relative velocity~$v_{2/1}$ of two UR particles emitted at the same event and in the same direction is constant during their travel}.

This statement---which is the physical content of the conjectures (5) in \FGMV---means that tidal forces (i.e. curvature effects) are negligible in this specific situation, even when the particles travel over cosmological distances. This is a priori surprising. Indeed, in the comparable situation of two UR particles, such as photons, emitted simultaneously but in slightly different directions, curvature is absolutely non-negligible, as it is responsible for all gravitational lensing phenomena.

In the next section, I propose a geometrical solution to this paradox, showing that for UR particles, \emph{longitudinal} curvature is effectively much smaller than \emph{transversal} curvature.

%%%%%%%%%%%%%%%%%%%%%%%%%%%
\section{Tidal forces and ultra-relativistic particles}
\label{sec:tidal_forces_UR}
%%%%%%%%%%%%%%%%%%%%%%%%%%%

\subsection{Geodesic deviation equation}

Because the particles $P_i$ are freely falling and very close to each other, it is reasonable to consider their worldlines~$\wl_i$ as infinitesimally separated timelike geodesics. Let us parametrize them by their own proper time $x_1^\mu(\tau), x^\mu_2(\tau)$, and introduce the separation vector~$\vect{\xi}$ defined by $\xi^\mu(\tau) \define x_2^\mu(\tau) - x_1^\mu(\tau)$. This vector is orthogonal to the geodesics, in the sense that $\xi^\mu u_\mu=0$, where $\vect{u}$ denotes the four-velocity of one of the particles. Physically speaking, $\vect{\xi}$ represents the spatial separation of the particles in their rest frame; its norm~$\xi^\mu \xi_\mu$ is thus nothing but the $\ell^2$ introduced in the previous section.

The evolution of $\vect{\xi}$ with the particles' proper time~$\tau$ is given by the geodesic deviation equation~\cite{2004rtmb.book.....P}
\begin{equation}
\Ddf[2]{\xi^\mu}{\tau} = -R\indices{^\mu_\nu_\rho_\sigma} u^\nu \xi^\rho u^\sigma,
\end{equation}
where $R\indices{^\mu_\nu_\rho_\sigma}$ are the components of the Riemann curvature tensor. This equation thus describes how curvature affects the relative motion of $P_1$ and $P_2$, which is precisely what we are worried about here.

For the remainder of this subsection, it will be convenient to work in the frame of, e.g., $P_1$, by choosing a Fermi normal coordinate system~$(\tau,X^a)$ about $\wl_1$. The orthogonality between $\vect{\xi}$ and $\vect{u}$ then implies that $\xi^\tau=0$, while the evolution of the spatial components~$\xi^a$ is
\begin{equation}\label{eq:tidal_effects}
\ddf[2]{\xi^a}{\tau} = -R\indices{^a_\tau_b_\tau} \xi^b \define -\tidal^a_b \xi^b.
\end{equation}
In the above equation, I have replaced covariant derivatives with normal derivatives thanks to the properties of Fermi coordinates, and introduced the $3\times 3$ tidal matrix~$\tidal^a_b\define R\indices{^a_\tau_b_\tau}$. An order of magnitude for the impact of tidal forces can then be obtained by integrating eq.~\eqref{eq:tidal_effects} perturbatively
\begin{equation}\label{eq:variation_velocity}
\Delta \dot{\xi}^a \approx - \pa{ \int_{\tau\e{s}}^{\tau\e{o}} \tau \tidal^a_b \; \dd\tau } \dot{\xi}^b\e{s},
\end{equation}
where $\dot{\xi}^b\e{s}$ is the initial relative velocity of the particles (at the source event), and $\Delta \dot{\xi}^a$ is the total variation of this velocity during the particles' travel to the observer. The integral in eq.~\eqref{eq:variation_velocity} can also be reformulated in terms of an average tidal matrix
\begin{equation}
\overline{\tidal}_{ab} \define \frac{2}{\tau_1^2} \int_{\tau\e{s}}^{\tau\e{o}} \tau \tidal_{ab} \; \dd\tau ,
\end{equation}
where $\tau_1 \define \tau\e{o}-\tau\e{s}$ still denotes the travel time in $P_1$'s frame.

If $\vect{e}$ denotes the unit vector in the direction of the initial relative velocity between $P_1$ and $P_2$, with $\dot{\vect{\xi}}_\source = v_{2/1}\vect{e}$, then eq.~\eqref{eq:variation_velocity} tells us that the fractional variation of $v_{2/1}$ is
\begin{empheq}[box=\fbox]{equation}\label{eq:relative_velocity_change}
\frac{\Delta v_{2/1}}{v_{2/1}} \approx -\frac{1}{2}\,(\overline{\tidal}_{ab} e^a e^b) \, \tau^2_1.
\end{empheq}

\vspace*{0.1cm}

\noindent From this result we can already understand how tidal effects may be suppressed for UR particles: as their effect on~$v_{2/1}$ goes like $\tau_1^2$, time dilation implies that if the particles travel over a distance $D$, then $\tau_1\sim D/\gamma_1$. Compared to a couple of non-relativistic particles travelling over the same distance~$D$, tidal forces are therefore effectively suppressed by $\gamma_1^{-2}$. As we shall see in the next subsection, this does not contradict the existence of gravitational lensing phenomena, because of the \emph{anisotropic frame-dependence of $\tidal_{ab}$}.

\subsection{Lorentz-boosted curvature}

Consider two frames along the worldline~$\wl_1$ of $P_1$: the particle's frame materialized by a tetrad $\{\vect{e}_\alpha\}$, such that $\vect{e}_0=\vect{u}_1$; and a reference frame~$\{\vect{e}_\mu\}$ obtained by parallely transporting the observer's frame at $O_1$ along $\wl_1$. Again, for simplicity we choose their orientations so that the motion of the particle in the reference frame occurs along $\vect{e}_1$. In this case, both frames are simply related by a (constant) Lorentz transformation
\begin{equation}
\vect{e}_\alpha = \Lambda\indices{^\mu_\alpha} \vect{e}_\mu,
\end{equation}
with
\begin{equation}
[\Lambda\indices{^\mu_\alpha}] =
\begin{bmatrix}
\gamma_1 & -\gamma_1 v_1 & 0 & 0 \\
-\gamma_1 v_1 & \gamma_1 & 0 & 0 \\
0 & 0 & 1 & 0 \\
0 & 0 & 0 & 1
\end{bmatrix}.
\end{equation}
If the observer is non-relativistic, then the components of the Riemann tensor over the reference tetrad are on the order of the square root of the Kretschmann scalar $R^{\mu\nu\rho\sigma}R_{\mu\nu\rho\sigma}$. In terms of orders of magnitude, on the surface of a massive gravitating body of mass~$M$ and radius~$R$, this quantity goes like $0.2 \,(M/M_\odot) (R_\odot/R)^3 \U{AU}^{-2}$; in a cosmological context, it is of order $(H_0/c)^2 = 0.2\U{Gpc^{-2}}$, indicating from which length scales curvature effects start to be non-negligible.

%\begin{equation}
%\dd s^2 = -(1+2\Phi) \dd t^2 + (1-2\Phi) \delta_{ij} \dd x^i \dd x^j,
%\end{equation}
%%
%then the Riemann tensor is (at first order in $\Phi$)
%%
%\begin{equation}
%R_{\mu\nu\rho\sigma} = 2 \delta_{\mu[\rho} \partial_{\sigma]}\partial_\nu \Phi - 2 \delta_{\nu[\rho} \partial_{\sigma]}\partial_\mu \Phi 
%\end{equation}
%%
%so that $\tidal_0 \sim \partial^2\Phi$. 

In the particle's frame, however, the Riemann tensor reads
\begin{equation}
R_{\alpha\beta\gamma\delta} 
= \Lambda\indices{^\mu_\alpha} \Lambda\indices{^\nu_\beta} \Lambda\indices{^\rho_\gamma} \Lambda\indices{^\sigma_\delta} R_{\mu\nu\rho\sigma}
\end{equation}
from which one can deduce the expression of the tidal matrix in the particle's frame, $\tidal\UR_{ab}$, as a function of its counterpart in the reference frame, $\tidal\REF_{ab}$, and of other components of the Riemann tensor in the latter frame:
\begin{widetext}
\begin{equation}
[\tidal\UR_{ab}] =
\begin{bmatrix}
\tidal\REF_{xx} & \gamma_1\tidal\REF_{xy} - R_{txyx} & \gamma_1\tidal\REF_{xz} - R_{txzx} \\
\gamma_1\tidal\REF_{xy} - R_{txyx} & \gamma_1^2\pa{ \tidal\REF_{yy} - 2 v_1 R_{tyxy} + v_1^2 R_{xyxy} } & \gamma_1^2\pa{ \tidal\REF_{yz} + 2 v_1 R_{t(yz)x} + v_1^2 R_{xyxz} } \\
\gamma_1 \tidal\REF_{xz} - R_{txzx} & \gamma_1^2\pa{ \tidal\REF_{yz} - 2 v_1 R_{t(yz)x} + v_1^2 R_{xyxz} }  & \gamma_1^2\pa{ \tidal\REF_{zz} - 2 v_1 R_{tzxz} + v_1^2 R_{xzxz} },
\end{bmatrix}
\end{equation}
%
%\begin{equation}
%[\tidal_{ab}] =
%\begin{bmatrix}
%R_{txtx} & \gamma R_{txty} - R_{txyx} & \gamma R_{txtz} - R_{txzx} \\
%\gamma R_{txty} - R_{txyx} & \gamma^2\pa{ R_{tyty} - 2 v R_{tyxy} + v^2 R_{xyxy} } & \gamma^2\pa{ R_{tytz} - 2 v R_{t(y t z)} + v^2 R_{xyxz} } \\
%\gamma R_{txtz} - R_{txzx} & \gamma^2\pa{ R_{tytz} - 2 v R_{t(y t z)} + v^2 R_{xyxz} }  & \gamma^2\pa{ R_{tztz} - 2 v R_{tzxz} + v^2 R_{xzxz} }
%\end{bmatrix}
%\end{equation}
%
with the symmetrization convention~$T_{(\mu\nu)}\define(T_{\mu\nu}+T_{\nu\mu})/2$. For example, if the metric reads $\dd s^2 = -(1+2\Phi) \dd t^2 + (1-2\Phi) \delta_{ij} \dd x^i \dd x^j$ in the reference frame, with $\Phi\ll 1$ and slowly varying with time, then the Riemann tensor in this frame reads $R_{\mu\nu\rho\sigma}=2 \delta_{\mu[\rho}\partial_{\sigma]}\partial_\nu\Phi - 2 \delta_{\nu[\rho}\partial_{\sigma]}\partial_\mu\Phi$, so that  $\tidal\REF_{ab} \approx \partial_a \partial_b \Phi$. In the particle frame, we get
%
%\begin{equation}
%[\tidal\UR_{ab}] \approx
%\begin{bmatrix}
%\partial^2_x\Phi & \gamma\partial_x\partial_y\Phi & \gamma\partial_x\partial_z\Phi \\
%\gamma\partial_x\partial_y\Phi & \gamma^2(1+v^2) \partial^2_y\Phi + \gamma^2 v^2 \partial^2_x\Phi & \gamma^2 \partial_x\partial_y\Phi \\
%\gamma\partial_x\partial_z\Phi & \gamma^2 \partial_y\partial_z\Phi  & \gamma^2(1+v^2) \partial^2_z\Phi + \gamma^2 v^2 \partial^2_x\Phi
%\end{bmatrix},
%\end{equation}
\begin{equation}\label{eq:boosted_tidal}
[\tidal\UR_{ab}] \approx
\begin{bmatrix}
\partial^2_x\Phi & \gamma_1\partial_x\partial_y\Phi & \gamma_1\partial_x\partial_z\Phi \\
\gamma_1\partial_x\partial_y\Phi & \gamma_1^2 (2\partial^2_y\Phi + \partial^2_x\Phi) & \gamma_1^2 \partial_y\partial_z\Phi \\
\gamma_1\partial_x\partial_z\Phi & \gamma_1^2 \partial_y\partial_z\Phi  & \gamma_1^2 (2\partial^2_z\Phi + \partial^2_x\Phi)
\end{bmatrix}
\approx
\begin{bmatrix}
\tidal\REF_{xx} & \gamma_1\tidal\REF_{xy} & \gamma_1\tidal\REF_{xz} \\
\gamma_1\tidal\REF_{xy} & \gamma_1^2(2\tidal\REF_{yy} + \tidal\REF_{xx}) & \gamma_1^2  \tidal\REF_{yz} \\
\gamma_1\tidal\REF_{xz} & \gamma_1^2 \tidal\REF_{yz}  & \gamma_1^2(2\tidal\REF_{zz} + \tidal\REF_{xx})
\end{bmatrix}
,
\end{equation}
where we have used that $v_1\approx 1$.
\end{widetext}
We conclude that the \emph{longitudinal} component of the tidal matrix~$\tidal_{||} \define \tidal_{xx}$, i.e. the curvature experienced by the particles in the direction of their motion, is the same whatever their velocity. On the contrary, the \emph{transverse} terms~$\tidal_{\perp} \define \tidal_{yy}, \tidal_{zz}$ are enhanced by $\gamma_1^2$,
\begin{empheq}[box=\fbox]{equation}
\tidal\UR_{||} = \tidal\REF_{||} \qquad \tidal\UR_\perp \sim \gamma_1^2 \tidal\REF_\perp.
\end{empheq}

In the second relation above, the $\gamma_1^2$ factor thus compensates the $\gamma_1^{-2}$ coming from time dilation in the geodesic deviation equation. As a result, if initially the relative velocity of the two particles is orthogonal to the boost, then eq.~\eqref{eq:relative_velocity_change} yields
\begin{equation}
\pa{ \frac{\Delta v_{2/1}}{v_{2/1}} }_\perp \sim \gamma_1^2 \overline{\tidal}\REF_\perp \tau_1^2 \sim \overline{\tidal}\REF_\perp D^2,
\end{equation}
the net effect on the particles' motion is therefore comparable to the non-relativistic case. This is the reason why gravitational lensing actually occurs.\footnote{Note also from eq.~\eqref{eq:boosted_tidal} that tidal effects are also enhanced by a factor two. This is the reason why deflection of UR particles is twice larger than the one of non-relativistic particles.}

If, on the contrary, the relative velocity of the particles is aligned with their velocity with respect to the reference frame, then
\begin{equation}
\pa{\frac{\Delta v_{2/1}}{v_{2/1}}}_{||} \sim \overline{\tidal}\REF_{\\} \tau_1^2 \sim \frac{\overline{\tidal}\REF_{||} D^2}{\gamma_1^2}.
\end{equation}
In this case, tidal effects are thus really suppressed by time dilation. The relative velocity~$v_{2/1}$ of two UR particles following each other can therefore be considered constant over their travel, even on cosmological distances where~$\tidal\REF_{||} D^2 \sim 1$. This finally validates eq.~\eqref{eq:time_delay_meaning}, i.e. the time-delay formula~\eqref{eq:time_delay}.

\section{Conclusion}

In this article, I have proposed a physical interpretation for the expression of the observed time delay between two UR particles emitted at the same event, recently derived by ref.~\cite{2016PhLB..757..505F}. This formula indeed means that the velocity of one particle, in the rest frame of the other, is constant; in other words the impact of tidal forces on their relative motion is negligible. This surprising result can be explained by a very general mechanism of relativistic gravitation: the curvature effectively experienced by UR particles in the direction of their motion is effectively suppressed by a factor $\gamma^{-2}$ with respect to the curvature they experience orthogonally to their motion. This is the reason why time-delay phenomena in gravitational optics are always much smaller than deflection phenomena.

In practice, the trajectory and energy of UR particles is also affected by non-gravitational interactions with galactic and intergalactic matter. Such interactions should generate astrophysical corrections to eq.~\eqref{eq:time_delay}, which must be evaluated in order to determine to which extent UR time delays are a viable cosmological observable.

\acknowledgements
I thank Chris Clarkson for encouraging me to write this article; I also acknowledge Giuseppe Fanizza, Maurizio Gasperini, Giovanni Marozzi, Cyril Pitrou, Jean-Philippe Uzan, Leo Stodolsky, and Gabriele Veneziano for relevant comments on the draft. The financial assistance of the National Research Foundation (NRF) towards this research is hereby acknowledged. Opinions expressed and conclusions arrived at, are those of the author and are not necessarily to be attributed to the NRF.

\bibliography{the_bibliography}

\end{document}